\documentclass[twocolumn,english,final]{article}
\pdfoutput=1

\usepackage[utf8]{inputenc}
\usepackage[T1]{fontenc}
\usepackage{cite}
\usepackage[fleqn]{amsmath}
	\allowdisplaybreaks[1]
\usepackage{abstract}

\usepackage{amsfonts}
\usepackage{amssymb}
\usepackage{microtype}
\usepackage[toc]{appendix}
\usepackage{hyperref}
	\hypersetup{%
		pdfauthor={Giulia Maniccia and Giovanni Montani},%
		pdftitle={WKB approach to the gravity-matter dynamics: a cosmological implementation}%
	}
\usepackage{cleveref}
\newcommand{\eu}{\mathrm{e}}
\newcommand{\iu}{\mathrm{i}}

\begin{document}
	
\title{WKB approach to the gravity-matter dynamics: a cosmological implementation}

\author{%
	G. Maniccia~\textsuperscript{1,2}\thanks{\href{mailto:giulia.maniccia@uniroma1.it}{giulia.maniccia@uniroma1.it}} \hspace{1pt}
	and G. Montani \textsuperscript{1,3}\thanks{\href{mailto:giovanni.montani@enea.it}{giovanni.montani@enea.it}}\\%
	\small {}\textsuperscript{1}Physics Department, ``La Sapienza'' University of Rome, P.le A. Moro 5, 00185 Roma, Italy\\%
	\small {}\textsuperscript{2}INFN Section of Rome, Piazzale Aldo Moro 2, 00185 Roma, Italy\\%
	\small {}\textsuperscript{3}ENEA, C.R. Frascati (Rome), Via E.\ Fermi 45, 00044 Frascati (Roma), Italy\\%
}
	
\date{7th October 2021}

\twocolumn[

	\maketitle
	
	\begin{center}
		\vspace{-1cm}
		\textit{Proceeding of 16th Marcel Grossmann Meeting, July 2021\\Quantum Gravity Phenomenology Parallel Session}
	\end{center}
	
	\begin{onecolabstract}	
		The problem of time emerging in the canonical quantization procedure of gravity signals a necessity to properly define a relational time parameter. Previous approaches, which are here briefly discussed, make use of the dependence of the quantum system on semiclassical gravitational variables in order to define time. We show that such paths, despite the following studies, lead to a non-unitary evolution. We propose a different model for the quantization of the gravity-matter system, where the time parameter is defined via an additional term, i.e. the kinematical action, which acts as a clock for quantum matter. The procedure here used implements a Born-Oppenheimer-like separation of the system, which maintains covariance under the foliation of the gravitational background and keeps the correct classical limit of standard quantum field theory on a fixed background. It is shown with a WKB expansion that quantum gravity corrections to the matter dynamics arise at the next order of expansion, and such contributions are unitary, signalling a striking difference from previous proposals. Applications to a cosmological model are presented and the analogies of the kinematical term with an incoherent dust are briefly discussed.
		\vspace*{\baselineskip}
	\end{onecolabstract}

]

\saythanks	
	
\section{Introduction}\label{intro}
	A suitable quantum description of the gravity-matter system is necessary in order to describe the Universe and its early phases. An important requirement for this purpose is the quantization of the gravitational field, which has opened a debate leading to many approaches and outcomes. 
	The first candidate for this procedure is the Dirac prescription \cite{bib:dirac-lecturesOnQuantumMechanics},  which leads however to the problem of time \cite{bib:isham-1992}: the Universe wave function does not evolve on time, hence a probabilistic interpretation is not clear.
	For this reason a different clock must be introduced \cite{bib:deWitt-1967,bib:isham-1992}: essentially, we have to define a \emph{relational time} \cite{bib:rovelli-1991} in order to describe the quantum system dynamics, adopting a subpart of the system as a clock through carefully chosen boundary conditions. 
	
	A first step in this path was proposed in \cite{bib:vilenkin-1989}, by separating the system into a semiclassical component and the purely quantum sector due to their different energy scales, in analogy with the Born-Oppenheimer approximation\cite{bib:prasad-datta-1997}. The external time was introduced via the dependence of the quantum subsystem  on the semiclassical variables and, with a Wentzel-Kramer-Brillouin expansion~\cite{bib:landau-quantumMechanics} (WKB) in the Planck constant, the limit of standard quantum field theory on curved background was recovered at the first order, together with a positive-semidefinite probability density under certain conditions (see the original article \cite{bib:vilenkin-1989} for such discussion). 
	
	Other models \cite{bib:kiefer-1991,bib:bertoni-1996,bib:kiefer-2018}, which differ for the choice of expansion parameter and initial conditions, have been studied focusing on the next order of expansion. The work \cite{bib:kiefer-1991} in particular dealt with computing quantum gravity induced corrections to the quantum matter dynamics, finding that non-unitary terms emerge. The workaround \cite{bib:kiefer-2018}, proposed to avoid such troublesome feature, cannot be applied in many contexts due to the strict initial requirements (see discussion in \cite{bib:maniccia-2021}).
	The work \cite{bib:bertoni-1996} followed more closely a Born-Oppenheimer approach on the problem, in order to define a conserved probability density; also in this case, however, it can be shown via the WKB expansion that the Hamiltonian operator describing the system dynamics is not unitary at the next order~\cite{bib:maniccia-2021}.
	
	Here we shall present a different implementation of the gravity-matter system, that ensures the standard limit of quantum field theory on curved background, with an expansion parameter proportional to the squared Planck mass (symbolizing the energy scale of gravity). The time parameter will not be linked to the dependence of the quantum sector on the semiclassical variables, but instead arise from an additional Hamiltonian term, i.e. the \emph{kinematical action} \cite{bib:kuchar-1981}, that reinstates  covariance of the theory in Arnowitt-Deser-Misner (ADM) variables\cite{bib:arnowitt-1962} for assigned foliations. This term, together with the definition of the deformation vector, will allow us to obtain a functional Schr\"{o}dinger equation for the quantum matter sector and, at the next order, a unitary dynamics including quantum gravity corrections.
	
	An application will be shown for the case of the homogeneous free massless scalar field in a spatially flat Friedmann-Lemaitre-Robertson-Walker background, representing the inflaton field during the de Sitter phase; in this case such quantum gravity corrections indeed modify the Hamiltonian spectrum, with non-vanishing but small contributions due to the perturbative nature of the approach.
	
	An important analogy is also discussed, linking the introduction of the \emph{kinematical action} to the frame fixing procedure using Gaussian coordinates\cite{bib:kuchar-1991}, that emerge as a fluid term (more specifically, an incoherent dust) in the action, and showing how the two procedures are related.
	
	The structure of the paper is the following. In Sec.~\ref{WKBexp} we will briefly introduce the WKB expansion of the gravity-matter system, discussing the mentioned approaches in literature and their outcomes. In Sec.~\ref{Proposal} we propose a model constructing the time parameter from the kinematical action and show that its WKB expansion gives a unitary dynamics for the quantum matter sector at the order next to standard quantum field theory. In Sec.~\ref{Application} the procedure is applied to a case representing the homogeneous inflaton field, in order to analyze the outcome in a cosmological view. In Sec.~\ref{KucharTorre} the analogy of such construction with the frame-fixing procedure is presented. Section~\ref{Conclusions} contains the conclusive remarks.

	\section{Main features of the WKB expanded gravity-matter system}\label{WKBexp}
	The starting point for the study of quantum gravity effects is the quantization of the gravitational field. It is known that, following the Dirac prescription \cite{bib:dirac-lecturesOnQuantumMechanics}, one finds a time-independent Schr\"{o}dinger equation which describes no evolution in time: the equation, known as the Wheeler-DeWitt \cite{bib:deWitt-1967} equation, is the essence of the so-called problem of time \cite{bib:isham-1992}.
	Then, in order to overcome such problem, one has to define a relational time parameter\cite{bib:rovelli-1991}, meaning that a selected subsystem must be chosen as a new, different clock to describe the evolution.
	
	Let us briefly recall how this critical result emerges.
	We start with the foliation of space-time proposed by Arnowitt, Deser and Misner\cite{bib:arnowitt-1962}: we identify 3d hypersurfaces immersed in the 4d spacetime environment by using some parametric equations, and derive the induces properties of such hypersurfaces from the environment ones. We label the induced 3d metric with the tensor $h_{ij}$, and the normal vector field with $n^{\mu}$. Using these variables, one can compute the Hamiltonian formalism of the gravitational field, finding primary and secondary constraints on the wave function of the system \cite{bib:cianfrani-canonicalQuantumGravity}. It follows that the wave function depends only on the equivalence class of the induced geometries $\{h_{ij}\}$, and on the matter fields $\phi_a$ if present. Due to the superHamiltonian constraint, the action of the superHamiltonian on the system wave function must vanish:
	\begin{equation}
		\label{eq:WDWinizio}
		\begin{split}
		H \Psi =& \left( -\frac{2 \hbar^2 \kappa}{\sqrt{h}} \nabla^2_{g}- \frac{\sqrt{h}\, R^{(3)}}{2 \kappa} \right.\\
		 &\left. -\frac{\hbar^2}{2 \sqrt{h}} \nabla^2_m + u(h_{ij}, \phi_a) \right) \Psi = 0
		 \end{split}
	\end{equation}
	where $h$ refers to the determinant of the induced metric, $R^{(3)}$ to its curvature, $\kappa$ is the Einstein coefficient $\kappa = 8\pi G/c^3$ and $u(h_{ij}, \phi_a)$ is the total potential energy of the matter fields. The gradients $	\nabla_g$ and $\nabla_m$ indicate respectively the derivatives with respect to the gravitational and matter variables.
	
	This result is troublesome since it implies the problem of time~\cite{bib:isham-1992}, where the associated Schr\"{o}dinger equation states that the Universe wave function does not evolve in time at all. It follows that another time parameter \cite{bib:rovelli-1991} should be implemented in order to describe a non-trivial, meaningful dynamics. 
	
	A semiclassical approach can be implemented in this sense. By following the Wentzel-Kramer-Brillouin expansion \cite{bib:landau-quantumMechanics}, we write the system wave function as the complex exponential $\eu^{\iu S/ \hbar}$ and expand it with a chosen parameter, examining the dynamics order by order. The choice of parameter will determine the range of validity: it can be chosen as the Planck constant $\hbar$, or related to the inverse of the Einstein coefficient $\kappa$, i.e. proportional to the square of the reduced Planck mass (here labeled $m_{\mathrm{Pl}}$):
	\begin{equation}
		\label{eq:defM}
		M \equiv \frac{1}{4 c^2 \kappa} = \frac{c \,(m_{\mathrm{Pl}})^2}{4 \hbar} .
	\end{equation}
	Although different parameters can be  implemented, these two cases present interesting results and therefore will be discussed in this paper.
	
	The works under examination \cite{bib:vilenkin-1989,bib:kiefer-1991} differ from each other in few but fundamental starting hypotheses; let us briefly discuss them in the following.
	
	\subsection{Expansion in Planck constant}\label{Vilenkin}
	
	In the $ \hbar \rightarrow 0 $ expansion\cite{bib:vilenkin-1989}, the variables are separated into semiclassical $c$ and quantum ones $q$, so that Eq.~\eqref{eq:WDWinizio} reads
	\begin{equation}
		\left( - \hbar^2 \nabla_c^2 + U_c (c) - \hbar^2 \nabla_q^2 + U_q (c, q) \right) \Psi (c, q) = 0 ,
	\end{equation}
	where the operator $ H_c $ is obtained neglecting all the quantum variables and $U_c, U_q$ are respectively the semiclassical and quantum sector potentials.
	
	The first assumption states that the action of the quantum Hamiltonian on the system wave function $\Psi$ is small with respect to the semiclassical one:
	\begin{equation}
		\label{eq:smallness}
		\frac{\hat{H}_q \Psi}{\hat{H}_c \Psi} = \mathcal{O}(\hbar) .
	\end{equation}
	Secondly, it is assumed that the semiclassical and quantum subspaces are orthogonal; stronger requirements are needed in order to perform the expansion after the quantum mechanical order~$ \mathcal{O}(\hbar) $ (see discussion in \cite{bib:maniccia-2021}). 
	
	These assumptions allow a clean factorization of the action, which is implemented at each order of expansion after the first one:
	\begin{equation}
		\label{eq:SfactorizationVil}
		S_n = \sigma_n(c) + \eta_n(c,q) ,\quad n \ge 1 
	\end{equation}
	so that
	\begin{equation}
		\label{eq:SExpansionVil}
		S =  S_0 + P + Q
	\end{equation}
	with
	\begin{equation}
	\quad P(c) = \sum_{n = 1}^{\infty} \hbar^n \sigma_n, \quad Q(c,q) = \sum_{n = 1}^{\infty} \hbar^n \eta_n,
	\end{equation} 
	being the term $S_0$ at lowest order $\mathcal{O}(\hbar^0)$ purely classical. This factorization is reflected in the wave function:
	\begin{equation}
		\label{eq:PsiDecompositionV}
		\Psi(c,q) = \psi(c) \chi(c,q) = \eu^{\iu (S_0 + P) / \hbar} \, \eu^{\iu Q / \hbar}  , 
	\end{equation}
	where the background part $\psi(c)$ is assumed to satisfy the semiclassical part of the Wheeler-DeWitt equation:
	\begin{equation}
		\label{eq:VilBackground}
		\left( -\hbar^2 \nabla_c^2 + U_c \right) \psi(c) = 0  .
	\end{equation}
	Expanding up to order $\hbar$,  a functional Schr\"{o}dinger equation is recovered for the quantum subsystem $\chi_q$ in the background given by the semiclassical variables:
	\begin{equation}
		\label{eq:SchrVil}
		\iu \hbar \frac{\partial \chi_1}{\partial \tau} = \hat{H}_q \chi_1  ,
	\end{equation} 
	with the following definition of time derivative 
	\begin{equation}
		\label{eq:defTimeVil}
		\frac{\partial}{\partial \tau} = 2\, \nabla_c \,S_0 \cdot \nabla_c \, .
	\end{equation}
	This implementation describes a non-trivial dynamics (Eq.~\eqref{eq:SchrVil}) and allows the definition of a positive-semidefinite probability density, with suitable requirements on the background foliation, so that the standard interpretation of the wave function for a small subsystem of the Universe is recovered. 
	
	It shall be noted that in this process the semiclassical variables, from which the time parameter is constructed, have been essentially treated as classical ones. At the next order of expansion $\mathcal{O}(\hbar^2)$, not present in the original work, the quantum properties of the background should start to influence the quantum subsystem. By continuing the calculation \cite{bib:maniccia-2021}, it can be seen that some correction terms to the quantum dynamics emerge:
	\begin{equation}
		\label{eq:schrCorrectedVil}
		\iu \hbar \frac{\partial \chi_2}{\partial \tau} = \hat{H}_q \chi_2 - \left( 2\iu \hbar^2\, \nabla_c \sigma_1 \cdot \nabla_c + \hbar^2\, \nabla_c^2 \right) \chi_2 .
	\end{equation}
	It is straightforward to prove that these additional terms, which are attributed to the non-purely classical behavior of the background, are not unitary, thus presenting us a crucial point of discussion.

	\subsection{Expansion in squared Planck mass}\label{Kiefer}
	The work \cite{bib:kiefer-1991} uses the parameter $M$ defined in Eq.~\eqref{eq:defM} for the expansion. Being it representative of the Planck scale, this choice implies that the lowest order will only see gravity and not the matter sector, describing a vacuum Universe in the limit $ M \rightarrow \infty $. Eq.~\eqref{eq:WDWinizio} becomes then:
	\begin{equation}
		\left( - \frac{\hbar^2}{2 M} \nabla_g^2 + M V (g) + H_m \right) \Psi (g, m) = 0 
	\end{equation}
	where $H_m$ contains all the matter terms. 	The wave function is factorized in a similar matter (see Eqs.~\eqref{eq:SfactorizationVil},\eqref{eq:SExpansionVil}) into a gravitational function and a matter one, each expanded with the WKB method:
	\begin{equation}
		\label{PsiDecompositionwithM}
		\Psi (g,m) = \psi(g) \chi(m;g) = \eu^{\iu (M S_0 + P) / \hbar} \,\eu^{\iu Q / \hbar} \, ,
	\end{equation}
	where the action at the lowest order (pure gravity) is of order $M$.
	
	Similarly, a time-dependent non-trivial Schr\"{o}dinger equation (see Eq.~\eqref{eq:SchrVil}) for the matter sector is obtained, defining again time from the dependence on the semiclassical gravitational variables:
	\begin{equation}
		\label{eq:defTimeKief}
		\frac{\partial}{\partial \tau} = \nabla_g S_0 \cdot \nabla_g .
	\end{equation}
	The authors also investigate the next order of expansion, finding that the quantum matter dynamics is described by:
	\begin{equation}
		\label{eq:schrCorrectedKief}
		\iu \hbar \frac{\partial \chi_2}{\partial \tau} = \hat{H}_q \chi_2 - \left( \frac{\iu\hbar}{M}  \nabla_g \sigma_1 \cdot \nabla_g + \frac{\hbar^2}{2M} \nabla_g^2 \right) \chi_2 ,
	\end{equation}
	presenting non-unitary terms. 
	
	Cosmological implications of such a result were presented in \cite{bib:kiefer-2016}. A possible workaround for the issue developed by the authors in  \cite{bib:kiefer-2018}, which eliminated the non-Hermitian part of the matter Hamiltonian by suitable redefinitions of the wave functions in the product $ \Psi = \psi \chi $, is not applicable to many cases (see discussion in \cite{bib:maniccia-2021}) due to the basic assumptions implemented, thus leaving the problem of how to deal with non-unitarity.
	We stress that both definitions of time (see Eqs. \eqref{eq:defTimeVil}, \eqref{eq:defTimeKief}) make use of the dependence of the quantum subsystem on the background semiclassical variables. 
	
	It is essential to cite the work \cite{bib:bertoni-1996}, which applied a procedure more similar to the Born-Oppenheimer method, by subtracting to the initial Eq.~\eqref{eq:WDWinizio} its average over the  quantum function $\chi$. However, once WKB expanded in the $M$ parameter, with a similar time definition as in Eqs.~\eqref{eq:defTimeVil}, \eqref{eq:defTimeKief}, the total quantum Hamiltonian operator still presents a non-unitary morphology~\cite{bib:maniccia-2021}. 
	
	\section{A different definition of time via the kinematical action}\label{Proposal}
	We have seen that the previously mentioned definitions of relational time, using the dependence on the semiclassical variables, give a non-unitary matter Hamiltonian operator due to quantum gravitational corrections (Eqs.~\eqref{eq:schrCorrectedVil}, \eqref{eq:schrCorrectedKief}). This result seems a direct consequence of such construction, even with different expansion parameters (we refer for this point to Ssec.~4.1 of the mentioned article \cite{bib:maniccia-2021}).
	
	For this reason we here propose a different construction, which implements the kinematical action \cite{bib:kuchar-1981} into the theory as a clock for quantum matter.
	
	\subsection{The kinematical action}
	The kinematical action, first introduced in \cite{bib:kuchar-1981}, is an additional action term used to ensure some constraint equations of a quantum system, in analogy with a Lagrangian multiplier. 
	We remind that, in the Hamiltonian formalism, gravity presents both primary and secondary constraints \cite{bib:cianfrani-canonicalQuantumGravity}; the components of the deformation vector of the ADM splitting \cite{bib:arnowitt-1962}
	\begin{equation}
		\label{eq:defVector}
		N^{\mu} = \partial_t y^{\mu} = N n^{\mu} + N^i b_i^{\mu} 
	\end{equation}
	play the role of the Lagrangian multipliers for the secondary constraints.
	
	When a specific splitting is assigned, however, the components $N$ and $N^i$ still appear in the action of the quantum field theory as multipliers, but their initial meaning as components of the four-vector $N^{\mu}$ is not evident. To recover this, the kinematical action in ADM variables can be inserted:
	\begin{equation}
		\label{kinAction}
		S_{k} = \int d^4 x (p_{\mu} \partial_t y^{\mu} - N^{\mu} p_{\mu}) ,
	\end{equation}
	where $y^{\mu} = y^{\mu} (x^i; x^0)$ are the coordinates defining the parametric equations of the hypersurfaces, and $p^{\mu}$ are the conjugate momenta. Inserting this term into the action, some additional equations of motion appear by variations of $y^{\mu}$, $p_{\mu}$ and $N^{\mu}$: these equations ensure \cite{bib:kuchar-1981} that the momenta $p_{\mu}$ are trivial and recover the physical meaning of the deformation vector $N^{\mu}$ as in Eq.~\eqref{eq:defVector}.
	
	Additional contributions also emerge in the Hamiltonian formalism to the superHamiltonian and supermomentum functions:
	\begin{subequations}
		\label{Hkin}
		\begin{gather}
			\label{eq:additionalH}
			\mathcal{H}^{k} = n^{\mu} p_{\mu} \, ,\\
			\label{eq:additionalHi}
			\mathcal{H}_i^{k} = b_i^{\mu} p_{\mu} 
		\end{gather}
	\end{subequations}
	which are linear in the conjugate momenta, a feature which is crucial in order to construct the time parameter of the theory.

	\subsection{Basic hypotheses and conditions}
	Let us now start with a Born-Oppenheimer-like separation of the wave function:
	\begin{equation}
		\label{eq:Ansatz}
		\Psi ( h_{a}, \phi, y^{\mu} ) = \psi (h_{a}) \chi (\phi, y^{\mu} ; h_{a})
	\end{equation}
	where $\psi$ is the slow gravitational sector, and $\chi$ the fast quantum one. We are considering the kinematical action as a matter term, regarding its variables as fast. This choice ensures the correct limit at the order of standard quantum theory on curved background.
	
	As stated in~\cite{bib:vilenkin-1989}, since the matter fields live at an energy scale much lower than the Planck scale, we can reasonably assume that the slow function satisfies the semiclassical Wheeler-DeWitt equation, as in Eq.~\eqref{eq:VilBackground}. For the sake of generality, we will not work in the minisuperspace, so that supermomentum constraints are not automatically satisfied. With the same reasoning we assume that the slow function satisfies the gravitational supermomentum constraint:
	\begin{gather}
		\label{eq:gravsuperHconstr}
		\left[ -\frac{\hbar^2}{2M} \left( \nabla_g^2 + g \cdot \nabla_g\right) + MV \right] \psi = 0 \, , \\ 
		\label{eq:gravsuperMconstr}
		2 \iu \hbar \, h_{i} \textrm{D}\cdot \nabla_g \psi = 0 \, ,
	\end{gather}
	where the term $g \cdot \nabla_g \equiv g_a \frac{\delta}{\delta h_a}$ is included to consider a generic factor ordering of the derivative operators.
	By insertion of the kinematical action, the equations satisfied by the total wave function $\Psi$ are:
	\begin{equation}
	\label{eq:totalsuperHconstr}
		\begin{split}
		&\left[ -\frac{\hbar^2}{2M} \left( \nabla_g^2 + g \cdot \nabla_g \right) + M V -\hbar^2 \nabla_m^2 \right.\\
		&\left. \hphantom{==}\vphantom{\frac{2}{2M}} +U_m \right] \Psi = \iu \hbar \, n^{\mu} \frac{\delta}{\delta y^{\mu}} \Psi \, ,
		\end{split}
	\end{equation}
	\begin{equation}
		\label{eq:totalsuperMconstr}
		\left( 2h_{i}\,\textrm{D} \cdot \nabla_g - \partial_i \phi \cdot \nabla_m \right) \Psi = \iu\hbar \, b^{\mu}_i \frac{\delta}{\delta y^{\mu}} \Psi\, .
	\end{equation}
	
	We implement $M$, defined in Eq.~\eqref{eq:defM}, as the expansion parameter, so that the perturbative approach will be valid for fields with small associated energy with respect to the Planckian one. Formally, this property can be stated as the smallness of the ratio
	\begin{equation}
		\label{eq:BORatio}
		\frac{\hat{H} \chi (\phi, y^{\mu}; h_{a})}{\hat{H}\psi (h_{a})} = \mathcal{O}\left(\frac{1}{M}\right)
	\end{equation}
	and the adiabatic approximation also implies the smallness of the variation of the fast wave function with respect to the slow variables:
	\begin{equation}
		\label{eq:derivQuant}
		\frac{\delta}{\delta h_{a}} \chi(\phi, y^{\mu}; h_{a}) \simeq \mathcal{O} \left( \frac{1}{M} \right) .
	\end{equation}
	
	\subsection{Emerging dynamics for the gravity-matter system}
	The model can now be  expanded following the WKB method. At the first order $\mathcal{O}(M^1)$, it is clear that only gravitational terms survive, recreating the gravitational Hamilton-Jacobi equation
	\begin{equation}
		\label{eq:HJ}
		\frac{1}{2} \nabla_g S_0 \cdot \nabla_g S_0 + V = 0
	\end{equation} 
	which ensures the classical limit of General Relativity. This is an intrinsic property of the expansion due to the choice of parameter, which selects the Planck scale.
	
	Following the expansion, at $\mathcal{O}(M^0)$ both the gravitational and the matter sector are present. Making use of the gravitational constraints~\eqref{eq:gravsuperHconstr}, \eqref{eq:gravsuperMconstr}, the equation for the matter wave function at this order (here labelled $\chi_0$) is found:
	\begin{equation}
		\begin{split}
		\label{eq:SchrodingerF}
		\iu\hbar \frac{\delta}{\delta \tau}\chi_0 &\equiv  \iu\hbar \int_{\Sigma} d^3 x \left(N n^{\mu} + N^i b^{\mu}_i\right) \frac{\delta }{\delta y^{\mu}} \chi_0 \\
		& = \hat{H} \chi_0
		\end{split}
	\end{equation}
	where the operator $\hat{H}$ is the matter Hamiltonian up to this order, defined combining both the matter superHamiltonian and supermomentum functions:
	\begin{equation}
		\label{defHmatter}
		\hat{H} (\bullet) = \int_{\Sigma} d^3 x 	\left(N \hat{\mathcal{H}}^m + N^i \hat{\mathcal{H}}^m_i \right) (\bullet) \, .
	\end{equation}
	
	Eq.~\eqref{eq:SchrodingerF} has been rewritten as a functional Schr\"{o}dinger equation through an integration over the spatial hypersurfaces and an appropriate definition of the time derivative, using the definition of the deformation vector \eqref{eq:defVector}. The striking difference from previous works is that, in this case, time comes from the kinematical action which plays the role of clock in the theory, instead of the dependence from the slow gravitational variables.
	
	The next order of expansion $\mathcal{O}(\frac{1}{M})$ gives for the matter wave function $\chi$, now enclosing the contributions up to this order, the following:
	
	\begin{equation}
		\begin{split}
		\label{eqfinale}
		\iu\hbar \frac{\delta}{\delta \tau} \chi = \hat{H} \chi & + \int_{\Sigma} d^3 x \left[\vphantom{\frac{1}{2M}}  N \nabla_g S_0 \cdot \left(-\iu\hbar \nabla_g \chi\right)\right. \\
		& \left.-2 N^k h_{k} \text{D} \cdot \left( \frac{1}{\chi}(-\iu\hbar \nabla_g \chi ) \right) \chi \right].
		\end{split}
	\end{equation}
	
	It can be shown \cite{bib:maniccia-2021} that the additional terms in Eq.~\eqref{eqfinale} are unitary. This property, together with the smallness of such contributions coming from the initial adiabatic assumption \eqref{eq:derivQuant}, make this model an acceptable one for investigating quantum gravity effects on the matter dynamics.

	\section{Dynamics of a cosmological scalar field}\label{Application}
	
	Let us now consider an homogeneous Universe filled with a free massless homogeneous scalar field, and with a cosmological constant $\Lambda>0$, a setup similar to \cite{bib:kiefer-2016}. 
	This model reproduces the inflationary scheme during the de Sitter phase, since the kinetic term of the scalar field stands for the kinetic term of the inflaton one, which can become important towards the Planck scale, and the cosmological constant accounts for the almost constant potential during that phase.
	We now apply the model described in the previous section to this simple case in order to infer the effect of the corrections computed in Eq.~\eqref{eqfinale} to the Hamiltonian spectrum of the scalar matter field.

	The background is given by a Robertson-Walker metric:
	\begin{equation}
		ds^2 = N^2 dt^2 -a(t)^2\, (dx^2+dy^2+dz^2)
	\end{equation} 
	with Ricci scalar:
	\begin{equation}
		R = \frac{6}{c^2} \left( \frac{\ddot{a}}{a} + \frac{\dot{a}^2}{a^2}\right)
	\end{equation}
	where $a$ is the cosmic scale factor. The shift vector $N^i$ (see Eq.~\eqref{eq:defVector}) cannot be included in the metric since it would violate isotropy of the model.
	
	Since we are considering the homogeneous case, choosing the ADM foliation such that the normal vector field is $n^{\mu} = (1,0,0,0)$, it is easy to see that the only surviving contribution of the kinematical conjugate momenta is
	\begin{equation}
		\hat{p}_0 = \iu \hbar \frac{\delta}{\delta T}
	\end{equation}
	where the constructed time parameter, here labeled as $T$, coincides with the synchronous time \cite{bib:montani-2002} hence $N=1$.
	
	Following the steps in Sec.~\ref{Proposal}, the total wave function of the system will satisfy the gravitational superHamiltonian constraint~\eqref{eq:gravsuperHconstr} and the total superHamiltonian constraint~\eqref{eq:totalsuperHconstr}. The second one, after integration over a proper portion of space, i.e. the fiducial volume $V_0$, gives:
	
	\begin{equation}
		\begin{split}
		\label{eqWDWapplication}
		&(H_{grav} +H_{\phi}+H_{kin}) \Psi = \vphantom{\frac{1}{2}} \\
		&=  \left( \frac{1}{2V_0 a^3} \pi_{\phi}^2 -\frac{2\pi G c^2}{3V_0\, a} \pi_a^2 +\frac{V_0}{8\pi G} \,\Lambda a^3 +p_0 \right) \Psi
		\end{split}
	\end{equation} 
	and analogously for the gravitational constraint. The contribution of the supermomentum is cancelled due to the condition $N^i=0$.
	
	In order to proceed with the expansion, the Einstein constant has to be rewritten in terms of the chosen parameter $M$ defined in Eq.~\eqref{eq:defM}. Then, the system composed of the two constraints can be expanded order by order as presented in Sec.~\ref{Proposal}.
	
	The gravitational constraint equation gives solutions for the functions $S_0, P_1, P_2$ of the gravitational wave function up to order $\mathcal{O}(1/M)$, following the separation in Eq.~\eqref{PsiDecompositionwithM}. Implementing these solutions, the total constraint gives at order $\mathcal{O}(M^0)$ the ordinary Schr\"{o}dinger equation for the scalar field:
	\begin{equation}
		\label{eq:TimeDerivativeAppl}
		\iu\hbar \frac{\delta }{\delta T} \chi = \hat{H} \chi
	\end{equation}
	where we have taken $V_0=1$.
	
	The corrections emerge at the next order $\mathcal{O}(1/M)$. After plugging the gravitational solutions into the total constraint equation, we switch to the time parameter $\tau$ such that $d\tau = \frac{dT}{a^3}$ and we pass to the Fourier space, so that the dynamical equation for the scalar field becomes:
	\begin{equation}
		\label{eqWDWapplFourier}
		\iu\hbar \frac{\delta \tilde{\chi}}{\delta \tau} = -\frac{\hbar^2}{2} \frac{\delta^2 \tilde{\chi}}{\delta \phi^2} +\hbar \, \frac{k_a\, (-\tau)^{7/3}}{3 (3\Lambda)^{1/6}} \,\tilde{\chi} ,
	\end{equation}
	where the matter Hamiltonian (given by the standard kinetic term of the free field) is indeed corrected by the additional term related to $k_a$, which is the eigenvalue of the conjugate momentum $\pi_a$ of the scale factor $a$.
	
	The solution for the matter wave function is
	\begin{equation}
		\label{solutionAppl}
		\tilde{\chi} = e^{ -\iu \frac{\hbar p^2 }{2} \tau \,+ \iu \frac{k_a \, (-\tau)^{7/3}}{7(3\Lambda)^{1/6}}  } \, ,
	\end{equation}
	where $p$ is the conjugate momentum with respect to the scalar field $\phi$. 
	
	The solution \eqref{solutionAppl} can be used to construct a wave packet and infer the magnitude of the spectrum modifications.	
	It is important to stress that, passing from Eq.~\eqref{eqWDWapplication} to Eq.~\eqref{eqWDWapplFourier}, the momentum $k_a$ and hence the cosmic scale factor maintain their quantum nature: for this reason we do not insert the classical limit $a(T)$ of the background metric, but we leave $a$ as an intrinsic quantum variable.	
	We also remind that, due to the initial assumptions (see in particular Eq.~\eqref{eq:BORatio}), the range of validity of the  expansion  is $-\frac{1}{M} <k_a <\frac{1}{M}$.
	
	Numerical analysis of the corrected Hamiltonian spectrum has been computed by considering a Gaussian wave packet in $a, \phi$ constructed with the solution in Eq.~\eqref{solutionAppl}. The results of the analysis show small corrections as expected from the hypotheses of the theory, since the quantum contributions are considered of order $\mathcal{O}(1/M)$.
	
	\section{Analogy with the frame fixing procedure}\label{KucharTorre}
	
	An interesting observation should be made regarding the procedure we followed. In a later paper~\cite{bib:kuchar-1991}, the authors fix the reference frame of a gravity-matter system by inserting an additional term to the action, which resonates with the kinematical action insertion developed years before from one of the authors~\cite{bib:kuchar-1981}. We do not show the detailed calculations here, reporting only the key points of such procedure and why it is connected to this work. 
	
	The chosen coordinates for their procedure are the Gaussian ones $g^{00} =1$, $g^{0i} = 0$, implemented with Lagrange multipliers $\mathcal{M}$, $\mathcal{M}_i$ into the action with an additional term, that is written in a parametrized form as:
	\begin{equation}
		\begin{split}
		\label{eqfluidSParam}
		S^F = &\int_{\Sigma} d^4x \left[ -\frac{1}{2} \mathcal{M} \sqrt{-g} (g^{\alpha \beta} \partial_{\alpha} T \partial_{\beta} T-1) \right.\\
		&\left.\vphantom{\frac{1}{2}} +\mathcal{M}_i \sqrt{-g} (g^{\alpha \beta} \partial_{\alpha} T \partial_{\beta} X^i) \right] \, ,
		\end{split}
	\end{equation}
	where $X^i, T$ are the Gaussian coordinates and $x^{\alpha}$ are the new variables (``parameters") associated to the metric $g_{\alpha \beta}$, so that coordinate transformations on the $x^{\alpha}$ are allowed.
	
	The authors prove that such term~\eqref{eqfluidSParam}, corresponding to the choice of reference system, materializes into the theory as a fluid, more precisely as an incoherent dust with energy tensor:
	\begin{equation}
		T^{\alpha \beta} = \mathcal{M}\,U^{\alpha} \,U^{\beta} \qquad \text{with} \;\, U^{\alpha} = g^{\alpha \beta} \partial_{\beta} T 
	\end{equation} 
	and with positive-definite energy as long as $\mathcal{M} \geq 0$.
	
	Again, going to the order of quantum matter, a Schr\"odinger equation emerges with time coinciding with the parameter $T$ of the  Gaussian frame. This means that the emergent fluid plays the role of clock for quantum matter, in a way which is very similar to the kinematical action implementation.
	
	Starting from the action with the ``emergent fluid term", we can implement the same WKB expansion as explained in Sec.~\ref{Proposal}. We obtain that, in the minisuperspace, such procedure gives at the order $\mathcal{O}(1/M)$ quantum gravity corrections which are isomorphic to those of Eq.~\eqref{eqfinale}.
	
	This result has an immediate explanation. In the minisuperspace, one can choose the ADM foliation such that $n^{\mu} = (1,\vec{0})$, consequently the parametric equations in \eqref{kinAction} give $y^0 = T, \,y^i = X^i$; this means that their time derivative becomes
	\begin{equation}
		\label{eqSKinAsFluid}
		\partial_t y^{\mu} \rightarrow \partial_t y^0 = \dot{T} = \frac{\delta T}{\delta t} 
	\end{equation}
	
	It is clear that, in the minisuperspace and with a specific foliation, the kinematical action exactly reduces to the ``reference fluid" emerging from the Gaussian reference frame fixing of Eq.~\eqref{eqfluidSParam}. In this sense, the two implementations are equivalent.

	\section{Concluding remarks} \label{Conclusions}
	
	The semiclassical WKB expansions of the gravity-matter system presented in the works here cited \cite{bib:vilenkin-1989,bib:kiefer-1991,bib:kiefer-2018,bib:bertoni-1996} encounter some problems when the canonical quantization procedure is applied and a time parameter is recovered from the dependence of the quantum system on the semiclassical variables. We have shown that the expansion performed in \cite{bib:vilenkin-1989} with parameter $\hbar$ describes a non-unitary quantum matter dynamics at the next order, as seen in Eq.~\eqref{eq:schrCorrectedVil}. A similar result already emerged in the work \cite{bib:kiefer-1991}, see Eq.~\eqref{eq:schrCorrectedKief}, which was later modified with an \emph{ad-hoc} procedure~\cite{bib:kiefer-2018}; however the basic assumptions of that workaround cannot be satisfied in many cosmological models, thus requiring a different procedure. 
	
	Even dealing with an approach more similar to the Born-Oppenheimer one \cite{bib:bertoni-1996}, after the WKB expansion, it seems that non-unitary terms for the matter sector dynamics do not emerge, but the theory cannot be set by means of a general Hilbert product (see discussion in \cite{bib:maniccia-2021}).
	
	The previous results signal a necessity to define the relational time in a different way, not using the dependence from the gravitational variables, which have an intrinsic quantum component due to the WKB limit.	For this reason a different construction of the gravity-matter system has been implemented, where both the WKB expansion and the adiabatic approximation have been used, but with the core difference that time has been constructed from the  the kinematical action. The utility of such term, defined in Eq.~\eqref{kinAction}, in quantum field theory on a given background is to recover covariance under ADM foliation, since the deformation vector is assigned due to an \emph{a priori} choice of the reference system.
	
	Thanks to the additional contributions (see Eqs.~\eqref{eq:additionalH}-\eqref{eq:additionalHi}) coming from the kinematical action, we have recovered a dynamics described by linear constraints in the conjugate momenta (Eqs.~\eqref{eq:totalsuperHconstr},\eqref{eq:totalsuperMconstr}) and covariant under choice of an ADM foliation. 
	The momenta $p_{\mu}$ associated to the kinematical variables $y^{\mu}$, after the quantization procedure, are a strong candidate for the definition of a time parameter. 
	
	By performing the WKB expansion in the parameter $M$ defined in~\eqref{eq:defM}, we have obtained at the highest order the Hamilton-Jacobi equation~\eqref{eq:HJ} for gravity, which is equivalent to the Einstein equations in vacuum \cite{bib:cianfrani-canonicalQuantumGravity}. At the next order, with the definition in Eq.~\eqref{eq:SchrodingerF}, we have recovered at the zero-th order a functional Schr\"{o}dinger equation which corresponds standard quantum field theory. Finally, at order $\mathcal{O}(1/M)$, the corrections arising from quantum gravity to the quantum matter sector dynamics have been computed, showing that the additional terms in Eq.~\eqref{eqfinale} are unitary.
	Although the obtained corrections are small by construction, they suggest a new investigation tool to evaluate how a non-standard dynamics of quantum field theory behaves, when examining energies small with respect to the Planck scale. We have shown in Sec.~\ref{Application} that such modifications give a non-zero contribution to the Hamiltonian spectrum of the scalar matter field in a cosmological setting.
	
	We have shown in Sec.~\ref{KucharTorre} the comparison between the kinematical action insertion and the reference frame fixing procedure \cite{bib:kuchar-1991}. Setting the problem in the minisuperspace and choosing a specific ADM foliation, the parametric equations giving the kinematical variables directly correspond to the Gaussian coordinates of the frame fixing procedure. Then, it is clear that the kinematical action in this case is equivalent to the choice of the reference system, that clearly emerges as a fluid in the theory. This analogy provides a starting point for a deeper understanding and future analysis on the presented model.

	\bibliographystyle{unsrt}
	\bibliography{proc_arxiv}

\end{document}